\documentstyle[12pt,mes_eso_scul]{article}
\begin{document}

\begin{titlepage}
\title{The ESO-Sculptor Faint Galaxy Survey: Large-Scale Structure and 
Galaxy Populations at 0.1 $\simless$ z $\simless$ 0.5}
\author{Val\'erie de Lapparent, Gaspar Galaz, St\'ephane Arnouts\\
        {\it CNRS, Institut d'Astrophysique de Paris}\\ 
        Sandro Bardelli, Massimo Ramella\\ 
        {\it Osservatorio Astronomico di Trieste}}
\maketitle

\end{titlepage}

\section{Introduction}

We describe the current status of the ESO-Sculptor Survey.  The
observational goal has been to produce a new multi-color photometric
catalogue of galaxies in a region located near the southern galactic
pole, complemented by a spectroscopic survey. The primary scientific
objectives are (1) to map the spatial distribution of galaxies at
$z\simeq0.1-0.5$ and (2) to provide a database for studying the variations
in the spectro-photometric properties of distant galaxies as a
function of redshift and local environment.

The first clues towards the understanding of the matter distribution
in the Universe have been obtained by mapping the distribution of its
major light emitting components, the galaxies.  One of the main
properties of the galaxy distribution is the presence of structures at
nearly the largest scales examined (of order of 100 \hmpc\ with a
Hubble constant of \H0) \cite{oort}. The 3-dimensional maps provided
by redshift surveys of various regions of the sky have clearly
demonstrated the inhomogeneity of the galaxy distribution and have
emphasized the need for systematic redshift surveys over large volumes
of the universe.

The nearby galaxy distribution suggests a remarkable structure in
which galaxies cluster along sharp walls which delineate vast regions
with diameters between 10 and 50\hmpc\ devoid of bright galaxies, in a
cell-like pattern\cite{cfa-slice}. Gigantic structures such as the
``Great Wall'' have been detected and pose the problem of the largest
scale for the inhomogeneities\cite{great-wall}. The general
distribution has the topological properties of a ``sponge'', which
naturally arises from gaussian initial perturbations collapsing under
gravity. The nearby redshift maps\cite{cfa-slice}
have generated a renewed interest in mapping the large-scale structure
of the Universe in the early 1990's.  Several ambitious programs were
then initiated, and provide new maps which probe the distribution out
to distances of $\sim500$\hmpc. These maps contain many sheets and
voids, and confirm the sponge-like topology\cite{lcrs,esp}, with no
evidence of voids larger than $\sim100$\hmpc\cite{muenster}. It seems
that these surveys have reached the scale where the galaxy
distribution becomes homogeneous, but this requires further
quantitative studies.

The maps of the galaxy distribution raise several fundamental
questions of observational cosmology. Among them is the problem of the
missing mass detected in increasing amounts at larger and larger
scales\cite{missing}. If the limits of the nucleo-synthesis
predictions are to be met\cite{nucleo}, the required dark matter for
explaining the formation of large-scale structure must be for the most
part non-baryonic\cite{cdm}. In this picture, the galaxy formation
must be biased towards the densest peaks of the matter
distribution\cite{biasing}. Therefore the voids of the galaxy
distribution could be filled with dark --- and partly non-baryonic ---
matter. So far, all observational searches for baryonic matter within
these voids in the form of galactic-size systems have led to only rare
detections (see \cite{brosch,faint-void} and references therein),
largely insufficient for explaining a significant fraction of the
missing mass. The detection of the dark matter is therefore a crucial
requirement for validating the current scenario for large-scale
structure formation (the gravitational collapse of primordial
fluctuations).  Another challenge is to reconcile the size and
amplitude of the inhomogeneities in the galaxy distribution with the
high degree of isotropy of the microwave background
radiation\cite{cobe} (this also requires large amounts of dark
matter).

Mapping the inhomogeneities in the galaxy distribution allows to
obtain clues on the nature of the primordial fluctuations in the
matter density field, and therefore to better understand the
prevailing mechanisms in shaping the early universe.  At large scales,
constraints on the spectrum of primordial fluctuations can be derived
directly from the maps of the 3-dimensional galaxy
distribution.  On smaller scales ($\sim1$\hmpc), the
non-linear effects of gravitation complicate the derivation of such
constraints.  A study of the different galaxy populations and their
environment becomes necessary for a better insight into the formation
of structure on galactic size up to several \hmpc. Constraints on the
initial perturbations which led to the formation of structures on
these scales may be obtained if one can eventually make the link
between the mass function of the collapsing matter, the star formation
history of the galaxies, and the influence of their environment.  The
mass function is predicted by the theoretical models and is partly
constrained by the galaxy luminosity function\cite{mass-func}.  The
star formation history is tightly constrained by the observations
(spectral energy distribution and magnitude number
counts)\cite{madau}. One difficulty is to decouple the influence of
the local environment, to which galaxies are intimately related via
tidal interactions and mergers, from the large-scale segregation
effects resulting from the initial conditions \cite{segre}.

Detailed knowledge of both the large-scale clustering and the galaxy
populations in a 3-dimensional galaxy map therefore provides
invaluable information for studying the formation and evolution of
structure in the universe.  With the goal to address these issues, the
ESO-Sculptor survey of faint galaxies (ESS, hereafter) was initiated
in 1989. The program was granted the key-program status by ESO,
which provided the unique opportunity for obtaining a new complete
galaxy sample to both substantial depth and area on the sky. A large
amount of observing nights on the 3.6m and NTT was attributed and
allowed to complete the observations in the fall of 1995. We describe
the characteristics of the photometric and spectroscopic ESS samples
in \S 2 and 3 respectively, and we report the major results already
obtained in \S 4. In \S 5 we comment on the results and examine the
prospects for the ESS in the coming years.
 
\section{The ESS photometric sample}

The photometric data for the ESS\cite{ess-phot} was obtained from CCD
imaging of a continuous strip of $1.53\dg(R.A.)\times0.24\dg(DEC.)
\simeq 0.37$ deg$^2$ in the Sculptor constellation (\env\ 0\hour
21\min, $\sim -30$\deg\ in J2000 coordinates), 17\deg\ away from the
Southern Galactic Pole.  The typical exposure times at the NTT in the
$B$, $V$, $R$ filters were 25 mn, 20mn, and 15 mn
respectively. Whereas the $R$ and $V$ images have comparable depths,
the $B$ images are shallower because of the lower quantum efficiency
at short wavelengths of the CCDs used. The photometry for the survey
is obtained by a mosaic of \env\ 50 CCD frames in each filter
overlapping on the edges. The galaxy catalogues are complete to
$B=24.5$, $V=24$ and $R=23.5$. All objects are detected in the 3 bands
up to $R\sim21.5$. In the range $22<R<23.5$, the color completeness
drops to 65\%. Figure 1 shows the projected distribution on the sky
for the $R$ photometric sample.

The data were reduced within MIDAS on SUN and IBM workstations. The
majors steps in the data reduction were: bias subtraction,
flat-fielding using a ``super-flat'' obtained from median-filtering of
the target exposures themselves, co-addition of multiple exposures,
removal of cosmic events. Because the photometric data were obtained
during 10 different observing runs with changing telescopes,
instruments, filters, and detectors, the calibration of the data
required a thorough work of measurement of color coefficients and
magnitude zero-points.  These were determined simultaneously with an
iterative method. Listed in Table 1 are the mean zero-points and the
colors coefficients for the various instrumental set-ups used at the
3.6m (EFOSC1) in 1989--1990, and at the NTT (EMMI) from 1990 to 1995
(see \cite{ess-phot} for further details). These allow to convert the
observed CCD magnitudes into the standard Johnson-Cousins $BVR$
system.

To obtain a homogeneous photometry, an algorithm for comparing and
adjusting the measured magnitudes in the overlaps of neighboring CCD
frames was used\cite{ess-phot}. This technique reduces by a factor of
3 the systematic deviations in zero-point between individual CCD
frames.  The resulting \rms internal uncertainties in our photometry are
0.04 mag in the 3 filters.  The galaxy catalogue was obtained by
analyzing the full CCD data with Sextractor \cite{sex}. Note that the
ESO-Sculptor data played a major role in testing and improving the
performances of the Sextractor package for faint galaxy
photometry. Because of its excellent performances, this software is
now widely used by the astronomical community. The final ESS
photometric catalogue was produced after identification of the
multiple detections of the same objects from different images, and
determination of the adopted objects parameters. This catalogue
provides aperture magnitudes in the standard $B$(Johnson),
$V$(Johnson) and $R$(Cousins) filters, astrometric positions to 0.2
arcsec, and morphological parameters for $\sim 9500$, $\sim12150$, and
$\sim13000$ galaxies respectively \cite{ess-phot,these-arnouts}. As a
by-product, the ESS also provide the 3-color photometry of 2143 stars
to $B=24.5$. These are useful for constraining the models of galactic
structure. 

The ESO-Sculptor galaxy number-counts in the 3 bands, and the color
distributions are in good agreement with the results for other
existing samples\cite{ess-phot} The faint number counts show the
well-known excess over no-evolution models, in all 3 bands
\cite{met-nat}. The faintest galaxies also exhibit a color evolution,
characterized by a blueing trend at $R>22$ in the $B-V$ color. These
effects are often interpreted as an increase in the star formation
rate with look-back time \cite{met-nat}.  The angular 2-point
correlation function also shows evidence for significant evolution at
$R\simeq23$ which could be related in origin to the excess of faint
galaxies in the blue number-counts \cite{these-arnouts}.  Whereas the
change in amplitude at fainter magnitudes was already detected in
several other samples, a change in the slope is also detected thanks
to the increased survey area compared to the previous studies. This
analysis of the angular 2-point correlation function might provide new
clues on the nature and evolution of galaxies at faint
magnitudes\cite{these-arnouts}.  Adjustment of the number-counts,
color distributions, and variations in the slope and amplitude of the
two-point correlation function for the ESS provide useful constraints
for the models of galaxy evolution, and might allow to discriminate
among the different scenarios (pure luminosity evolution, dwarf galaxy
component etc...) \cite{met-nat}.

\section{The ESS spectroscopic sample} 

The spectroscopic catalogue provides the flux-calibrated spectra of
the complete sub-sample of $\sim$700 galaxies with $R_c\le20.5$ using
multi-slit spectroscopy. At this depth, there are 1.3 galaxy per
sq. arcmin in the survey, and the typical multi-slit masks at the NTT
contain \env\ 30 slits. The spectra are reduced using semi-automatic
MIDAS procedures which were specifically designed for these data and
which guarantee a homogeneous and systematic treatment of the numerous
CCD exposures and extracted spectra.  The major steps in the
spectroscopic reduction are: 2-D correction for vignetting of field;
cosmic events removal by comparison of multiple exposures;
flat-fielding to correct for pixel-to-pixel variations, for variations
in the slits transmission, and for fringes; long-slit wavelength
calibration using the context LONG of MIDAS; sky subtraction with a
wavelength-dependent fit of the sky flux along the slit and by
interpolation at the position of the object; optimal extraction of
objects by profile weighting; flux calibration using
spectrophotometric standards; cross-correlation of the resulting
spectra with galaxy templates for redshift measurement and error
estimation.

The delicate extraction of spectra of extended objects with an
integrated luminosity which represents only a fraction of the night
sky, as it is the case for the ESS, requires optimized techniques. As
a result, several of the high level commands have been re-written
within MIDAS, and allow better control of the reduction.  In
particular, the cross-correlation program was specifically written
for and tested on the ESS data. The essential steps are the continuum
subtraction, the filtering of the spectra, and the matching of the
rest-wavelength intervals of the observed spectrum and template. This
matching is performed by first estimating the object redshift based on
the cross-correlation with 6 templates representing the different
galaxy spectral types (E, S0, Sa, Sb, Sc, Irr) and obtained by
averaging several spectra from Kennicutt's sample \cite{kenn}.  The
spectra are then cross-correlated within their common rest-wavelength
intervals with a template of the bulge of M31 which provides
a reliable zero-point of the redshift scale \cite{ess-spec}. We
emphasize that we do {\it not} use the comparison of the
cross-correlation with the different galaxy types for determination of
the spectral types of the ESS galaxies: this technique is subject to
large errors because it is very sensitive to the noise and spurious
features in the spectra. Our spectral classification method is
described in \S 4.2 below.

The full MIDAS routine guarantees a well-controlled and user-friendly
reduction of the numerous images containing simultaneous spectra of 10
(with EFOSC1) to 30 (withEMMI) galaxies.  The average slit lengths vary between
$\sim 10$ and $\sim 30$ arcsec and the slit width is in the range
1.3--1.8 arcsec.  The dispersion is 230 \AA/mm. The resolution of the
resulting spectra varies from 20 \AA\ (EFOSC1) to 10 \AA\ (EMMI) (\env 110
spectra where obtained with EFOSC1 at the 3.6m, the rest with EMMI at
the NTT).  The resulting wavelength coverage is 4300--7000 \AA\ with EFOSC1
and 3500-9000 \AA\ with EMMI. The signal-to-noise ratio of the
spectra varies in the range 4--40, with 75\% in the spectra in the
range 10--30. The resulting \rms uncertainties errors in the redshifts
are in the range 0.0002 to 0.0005 (\ie \env 60 to 150\kms; this
corresponds to a spatial scale of \env\ 1\hmpc, which is small
compared to the size of the large-scale structures). 
For each galaxy, the rest-wavelength
interval results from the combination of the grism dispersion, the
position of the object within the multi-aperture mask, and the
object redshift.  The majority of the EFOSC1 spectra have
rest-wavelength intervals in the range 3300--4700 \AA\, and the EMMI
spectra in the range 3300--5700 \AA.  Therefore, only few of emission-line 
galaxies have H$\alpha$ within the observed wavelength range.

Typical spectra of the ESS in the interval 3700--5350 \AA\ are shown
in Figure 2 (black lines).  The prominent H (3968.5 \AA) and K (3933.7
\AA) lines of Ca II and the G molecular band of CH (4304.4 \AA) are
detected in most spectra with no emission lines. A large part of the
contribution to the cross-correlation peak originates from these
lines.  A large fraction (48\%) of the galaxies in the sample have
emission lines \cite{ess-class}. These are essentially [OII] 3727 \AA,
H$\beta$ 4851 \AA\ and [OIII] 4958.9 \AA\ and 5006 \AA. For
objects with emission lines, the final redshift is obtained by the
weighted mean of the cross-correlation redshift with the emission
redshift derived by gaussian fits to the emission lines.  Among the
277 galaxies for which the spectral classification has already been
performed (see \S 4.2), 4 are most likely HII galaxies (based on
diagnostic diagrams using line ratios), and only one galaxy is a
Seyfert 2 \cite{ess-class}. This is in marked disagreement with the
significantly larger fraction of active galaxies found in the
Canada-France Redshift Survey at $z\le0.3$ \cite{cfrs-tresse}.

A large fraction of the ESS spectra (\env 3/4) were observed in
spectro-photometric weather conditions. Note that optimization of the
multi-slit spectroscopic observations does not allow to minimize the
flux losses by adjusting the slit width and slit orientation (to
correct for atmospheric refraction): the slit width is fixed and
varies from 1.3 to 1.8 arcsec for the survey, and the slit/mask
orientation is chosen as to globally maximize the spacings between the
objects perpendicular to the dispersion direction.  However, at the
redshift of the ESS galaxies ($z>0.1$), the slits used contain $>95$\%
of the disk and bulge emission of a typical face-on spiral galaxy
($\sim20$ kpc in diameter). Therefore, the aperture and orientation
bias which affect the nearby redshift surveys as well as the
intermediate distance multi-fiber surveys, are small for the ESS.
Comparison of multiple spectroscopic observations for a sub-sample of
40 spectra shows that for those taken in spectro-photometric
conditions, the total external error is of order of 7\%
pixel-to-pixel. Because the only existing flux-calibrated samples of
galaxies are nearby samples of several tens of spectra \cite{kenn},
the ESS spectra provide a unique database of galaxy spectra
representative of the galaxy populations in a significant spatial
volume.  These spectra will be useful for constraining at low redshift
the models of spectrophotometric evolution of galaxies, which is an
essential step for making reliable predictions of galaxy evolution at
high redshift.

\section{Current results}

\subsection{Large-scale structure}

The ESS allows for the first time to map in detail the large-scale
clustering at $z\simless 0.5$ (nearly 10 times deeper than the nearby
maps of the galaxy distribution\cite{cfa-slice}).  Figure 3 shows the
spatial distribution for \env 400 galaxies of the ESS in 3 portions
because of the long line-of-sight (the right cone shows the full
redshift range for the ESS). These maps reveal a highly structured
distribution which closely resembles that seen in the nearby surveys:
the distribution is characterized by an alternation of sharp walls
which are spatially extended across the sky, with voids with typical
diameters of $20-60$\hmpc\cite{ess-lss}. The ESS maps suggest that the
cell-like galaxy clustering seen in the shallower redshift surveys
extends to $z\simeq0.5$.  Several statistical analyses of the data are
in preparation (two-point correlation function, power-spectrum, genus,
etc...)  and will provide a characterization of the large-scale
clustering in the ESS for comparison with the nearby galaxy
distribution. The long line-of-sight sampled by the ESS might also
provide new constraints on the galaxy clustering on scales of
$\simgreat$100\hmpc, which are poorly sampled by the shallower
surveys.

We emphasize that the large-scale structure pattern detected in the
ESS is nowhere as regular as in the redshift survey of Koo \etal
\cite{koo}, which suggests a periodic distribution of dense structures
with a separation of 128\hmpc.  This scale if well above the typical
size of the voids in both the shallower surveys
\cite{cfa-slice,lcrs,esp} and the ESS. This disagreement can
be partly explained by the narrow beam size of the Koo \etal probes (\env\
5\hmpc\ at $z\sim0.4$) which makes them sensitive to small-scale
clustering and might cause an overestimation of both the density contrast
of the walls and the size of the voids.  Moreover, the sparse
distribution of the narrow probes of Koo \etal over the sky (only a
few percent of the galaxies in the survey region have a measured
redshift) makes it difficult to establish a relationship between the
detected peaks and the network of sheets and voids.  In contrast, the
ESS was designed to sample efficiently structures similar to those
found in the nearby surveys: at the median redshift of 0.3, the
redshift survey probes an area of $4\times15$\hmpcs, sufficient for
detecting most sheets with a surface density comparable to that for
the ``Great Wall''\cite{great-wall}.  Note that the ESS lies 7\deg\
away from the Koo \etal survey on the sky. At $z\simeq0.3$ this
corresponds to 90\hmpc\ (with $q_0=0.5$). Comparison of the 2 surveys
shows no obvious correlation in the occurrence of the over-dense
structures along the line-of-sight, which might be difficult to
reconcile with a typical clustering scale of 128\hmpc. Further
investigations along this line using simulated distributions are in
progress.

\subsection{Spectral classification}

In the ESS, galaxies have diameters $\le 12$ arcsec. As a result, the
disk and spiral arms are poorly visible. Any attempt for a
morphological classification would thus be largely approximative, and
could only be limited to the presence of the main features (disk,
bulge, spiral arms, signs of interaction), and restricted to the
closest objects (to $z\le 0.2$; at larger distances, the diameter of
the objects becomes too small --- $<9$ arcsec --- for any usable
classification). Even from high resolution images, the morphological
classification is dependent on the filter used for the imaging, and
different filters show different stellar components with varying
morphologies\cite{uv-morph}.  The Spectral Energy Distributions (SED hereafter)
are a useful alternative approach for characterizing the galaxy
populations.  The SEDs measure quantitatively the
relative contributions of the underlying stellar components and
constrain the gas content and average metallicity. The spectral
classification thus provides a physical sequence which can be
interpreted in terms of evolution of the stellar components, and allows
to trace back the episodes of stellar formation.  For deriving a
robust classification for the ESS, we have used the Principal
Component Analysis (PCA hereafter) \cite{murtagh}.  This technique is
un-supervised in the sense that it does not rely on the use of a set of
galaxy templates. It provides an objective study of the systematic and
non-systematic trends of the sample, and has the advantage of being
poorly sensitive to the noise level in each spectrum. Moreover, the
resulting spectral classification is strongly correlated with the
Hubble morphological type\cite{connolly,sodre,ess-class}.

Application of the PCA to the ESS allows to re-write each spectrum as a
linear combination of a reduced number of parameters and vectors (3 in
this case), and which accounts for 98\% of the total flux in each
spectrum. The spectral type of the galaxies can be written in terms of
2 independent parameters $\delta$, $\theta$ which respectively measure
the position of the spectra along the sequence of spectral types, and
the deviation from the sequence originating from either peculiar
continua and/or strong emission line.  The two parameters are in
addition correlated (late-types tend to have stronger emission lines).
Figure 4 shows the $\delta$--$\theta$ sequence for 277 galaxies of the
ESS sample. Figure 5 shows the first 3 principal components obtained
from these data set (PC1, PC2, and PC3). The first 2 principal
components account for the red and blue stellar populations in the
observed galaxies, and their relative contributions to each galaxy
spectrum define its position along the PCA sequence (measured by
$\delta$); the 3rd component determines the emission line contribution
(measured by $\theta$). It was already known that the colors of
galaxies can be described by a linear combination of stellar colors
(namely types AV and M0III\cite{aaronson}). The interest of the PCA is
to provide a more detailed demonstration of this effect over a large
sample of galaxy spectra.

The continuous PCA spectral sequence obtained for the ESS can be
binned to provide the corresponding fractions of the different Hubble
morphological types (the correspondence is made by using spectra of
galaxies with known morphology \cite{kenn}).  In this manner, we find
that the ESS contains 17\% of E, 9\% of S0, 15\% of Sa, 32\% of Sb,
24\% of Sc, and 3\% of Sm/Im (see Figure 4).  The type fractions show
no significant variations with redshift over the redshift range
$0.1<z<0.5$, and are in good agreement with those found from other
surveys to smaller or comparable depth (see
\cite{ess-class,these-galaz}).  We find systematic variations in the
main spectral features (equivalent width of the [OII], [OIII] and
H$_\beta$ emission lines; height of the 4000 \AA\ break; slope of the
continuum) with PCA spectral type, which illustrates the efficiency of
the PCA technique for performing a physically meaningful spectral
classification.

The PCA spectral classification has many advantages over other
classification methods. It first provides a {\it continuous}
classification in a 2-parameter space, which allows quantitative
analyses of the sample properties as a function of spectral type. In
particular, it will allow an unprecedented measurement of the
morphology-density relation \cite{dressler} at large distance. The PCA
also provides a convenient filtering technique: the reconstructed
spectra (with 3 components) are inherently ``noise-free'' because the
principal components are derived from a large sample of spectra. For
the ESS, the reconstructed spectra have a signal-to-noise in the range
35--80 (to be compared with the range of 4--40 for the S/N of observed
spectra). The ESS spectra are reconstructed from the principal
components of Figure 5 as a linear combination $\alpha_1 {\bf PC1}+
\alpha_2 {\bf PC2}+\alpha_3 {\bf PC3}$ (with $0.92<\alpha_1<1$,
$-0.2<\alpha_2<0.3$ $-0.05<\alpha_3<0.15$).  Figure 2 shows the
reconstructed spectra (in red) versus the observed spectra (in black)
for 27 ESS galaxies.  Reconstruction of noise-free spectra can be
especially useful for comparing the ESS spectra with synthetic
templates obtained from models based on stellar population
synthesis\cite{pegase}.

The spectral classification for the ESS allows to derive accurate
cosmological K-corrections for the various photometric bands, which
correct for the ``blue-shift'' of the {\it observed} filter bands with
respect to the {\it rest-frame} spectra. The crucial step is the
extrapolation of the observed spectra in the rest-frame $B$ band,
which at $z\sim0.5$ corresponds to a $U$ filter. For this, we use the
multi-spectral model PEGASE developed at IAP by B. Rocca-Volmerange
\etal \cite{pegase} which provides galaxy SEDs from the UV to the
infra-red. These represent a good match to the ESS spectra in the
optical range 3700--5250 \AA.  We can then derive analytical relations
between galaxy spectral type as measured by $\delta$, redshift and
K-correction\cite{these-galaz}. In turn, the K-corrections provide
absolute magnitudes for the galaxies in the rest-frame filter
bands. Figure 6 shows the observed $B-R$ colors for the ESS spectra as
a function of redshift and spectral type, and the intrinsic colors
after application of the K-corrections. These diagrams show that the
effect of the K-corrections on the object colors is significant over
the redshift range for the ESS, and is strongly correlated with
spectral type. Note however that the large dispersion in the observed
$B-R$ colors makes any attempt to determine spectral types from
color-redshift diagrams subject to large uncertainties. This approach
is often used for determination of the galaxy types. 

\subsection{Luminosity function}

In addition to the physical information which it provides, the galaxy
luminosity function is indispensable for any statistical study of an
apparent magnitude-limited survey. However, this function is poorly
known so far due the the limited samples adequate for its measurement.
The existing luminosity functions measured at $z\le0.2$ from surveys
with typically $\sim10^4$ galaxies give variables results, which are a
function of the selection criteria for the samples\cite{these-galaz}.
The ``local'' luminosity function ($z\sim0.03$) \cite{marzke-lf2} is
likely to be affected by errors in the magnitude measurements in the
Zwicky catalogue.  Luminosity functions based on digitized plates can
be biased by the non-linearity of the photographic emulsion at bright
magnitudes \cite{bertin}, and by incompleteness at faint magnitudes.
The recent Las Campanas Redshift Survey\cite{lcrs} has the advantage
of providing the largest CCD galaxy catalogue for measurement of the
luminosity function at $z\sim0.2$ (see \cite{lcrs-lf} and references
therein). However, this spectroscopic sample is affected by selection
effects such as variable sampling over the sky, and a systematic
under-sampling of low-surface brightness galaxies, which might be
responsible for the shallow faint-end slope.

The luminosity functions for different galaxy types show variations in
their faint end slopes\cite{lcrs-lf,marzke-lf2}; in some cases, the
variation is closely related to the presence of emission lines in the
galaxy spectra \cite{esp-lf}. In addition, deep redshift surveys
suggest that the slope of the late-type luminosity function evolves
significantly at
$z\ge0.5$\cite{autofib-lf2,cnoc1-lf,cfrs-lf}. Although the different
deep surveys agree to detect an evolution in the luminosity density by
a factor of nearly 2 between $z\sim0$ et $z\ge0.5$
\cite{cfrs-lf,hdf}, the differences in the results emphasize the need for a
confrontation with new catalogues.

In Figure 7, we show the B and R luminosity function for 327 galaxies
from the ESS sample calculated using an ``inhomogeneity-independent''
method (see \cite{these-galaz} for details).  The 2 luminosity functions
are in good agreement with the results for the CNOC1 survey
\cite{cnoc1-lf} using similar filters and an analogous observational
setup (multi-slit spectroscopy at the Canada-France-Hawaii telescope),
and which probes the galaxy distribution to similar depth as the ESS.
A more detailed study using the full ESS sample is in course.

\section{Conclusions and prospects}

The ESS demonstrates the interest of a deep fully-sampled pencil-beam
survey for probing through and identifying numerous large-scale
structures along the line-of-sight.  The survey confirms that the
nearby properties of the large-scale clustering extend to
$z\simless0.5$, namely a cell-like structure of sharp walls
alternating with voids of order of 20--60\hmpc\ in diameters
\cite{ess-lss}. Measures of the power-spectrum at scales
$\simgreat100$\hmpc\ and of the topological properties of the detected
structures will provide useful constraints on the nature of the
primordial fluctuations which led to the observed large-scale
clustering.  The ESS also has the potential for uncovering very large
structures exceeding the extent of the shallower surveys. Some
marginal evidence for the presence of an extended under-density in the
redshift range 0.3--0.4 is under close examination.

The ESS photometric survey is the largest CCD multi-color survey of
galaxies, and allows to confirm with tighter error bars the previous
analyses of galaxy number counts and color distribution at $B<24.5$
based on smaller areas and/or fewer bands \cite{ess-phot}.  The ESS
galaxy spectra provide an unique database for adjustment of the models
of spectrophotometric evolution of galaxies at ``low''
redshift\cite{pegase}.  Calibration of these models on the ESS data,
via the proportions of the different galaxy types and their luminosity
functions will allow to obtain better predictions of galaxy evolution
at $z>1$. These predictions could be tested on the full photometric
sample, which extends significantly deeper than the spectroscopic
sample, using the 2-point angular correlation function
\cite{these-arnouts}.  Photometric redshifts techniques\cite{hdf},
which would require the acquisition of $U$ band photometry, would
provide additional constraints, and are under consideration.

The ESS spectroscopic sample is being complemented by a similar
spectroscopic survey being performed in the northern hemisphere using
the CFH Telescope.  The $\sim1700$ galaxies contained in the two
samples, and their homogeneous spectral classifications and absolute
magnitude determinations provided by the Principal Component Analysis
\cite{ess-class} will provide a new measure of the luminosity function
as a function of spectral class at the intermediate redshifts $0.1\le
z\le0.5$. The difficulty to interpret the galaxy number-counts from
the ``Hubble Deep Field'' \cite{hdf} emphasizes the need for a better
determination of the ``local'' luminosity function per galaxy type.
Objective detection of the galaxy groups within the ESS is also in
progress and will allow a detailed study of the different galaxy
populations and their relationship with the environment, as measured
by the local galaxy density and the location within the large-scale
structure. In particular, the morphology-density
relation\cite{dressler}, and the existence of an analogous to the
Butcher-Oemler effect \cite{bo} for field galaxies are being
investigated.

The optical study of the ESS data is complemented by multi-wavelength
follow-up observations: IRAC2 on the ESO-2.2m telescope is used for
obtaining $K'$ imaging of a sub-sample of the ESS spectroscopic
sample; the same regions is scheduled for observations with ISO at
10$\mu$ and 90$\mu$ (in collaboration with B. Rocca-Volmerange); the
full ESS region has been observed with the VLA at 6 cm and 20 cm (in
collaboration with J. Roland and A. Lobanov). The optical-$K'$ colors
allow to identify the stellar populations from massive stars to old
giants. The $K'$-far-infra-red colors allow to separate the different
populations of grains, and indirectly allow to constrain the star
formation rate.  Determination of the luminosity functions per type in
the infra-red, are crucial for interpreting the infra-red galaxy
counts. These, in contrast to the optical counts, show no excess over
no-evolution models\cite{cowie}.  Finally the radio observations will
allow to study the correlation between the optical and radio
properties of the different galaxy populations in an-optically
selected sample (the opposite of the usual approach of making optical
follow-up observations of a radio-selected sample).

We are considering the extension of the ESS spectroscopic sample one
magnitude fainter using FORS1 at the VLT/UT1. This would provide the
redshifts for another \env\ 1000 galaxies to $z\simless 1$ for which
the photometry is already available. This survey would be useful for
obtaining a dense sampling of the large-scale structure at redshifts
where so far only individual structures are detected, usually in
association with quasars or radio-galaxies. The evolution of the
large-scale clustering at $z\sim1$ is expected to depend markedly on
the cosmological parameters\cite{white}. Any detection or absence of
evolution in the cell-like pattern with redshift would provide useful
constraints on the mean matter density in the Universe ($\Omega$). The
redshift extension of the ESS would also be useful for preparing the
definition of the large area surveys which will be done more
efficiently with VIRMOS.

Only one other systematic redshift survey is currently being performed
at the depth of the ESS: the CNOC2 program, with the goal to obtain
redshifts for 10,000 galaxies at $z\simless0.7$.  Answering the
question of the scale of homogeneity in the galaxy distribution, and
hence of the underlying matter distribution will nevertheless require
larger area redshift surveys than the ESS and CNOC2. In particular,
the ``Sloan Digital Sky Survey'' (SDSS) \cite{sdss} and the ``2dF''
project \cite{2dF} to map 1 million and 250,000 galaxies respectively
out to distances $z\sim0.2$ over large areas of the sky will both make
a tremendous improvement in the statistical analysis of the galaxy
distribution. The larger distances will be probed by the 2dF extension
(6,000 galaxies to $R\simeq21$) \cite{2dF} and the DEEP survey with
the Keck Telescope (15,000 galaxies with $B\simless24$) \cite{deep}.
Note that the well-known difficulties with the flux calibration of
multi-fiber spectroscopy, supplemented by the aperture bias at
$z\simless0.2$ (the fibers only sample the core of the galaxies) will
make any spectral classification of the SDSS and shallow 2dF survey
subject to a number of biases.  A unique survey for determination of
the galaxy luminosity function per type will be the 5-m ``Large Zenith
Telescope'' which will provide SEDs with photometric quality for
nearly one million galaxies to $z\sim1$ using a liquid-mirror
telescope and a multi-narrow-band imaging technique\cite{lzt}.  These
data will be essential for constraining the evolution of the galaxy
populations and of the large-scale clustering with look-back time. It
will also allow to simulate the biases in the SEDs obtained with the
multi-fibers surveys.  Although the ESS is not in proportion with
these large-area surveys to come in terms of survey volume, budget,
and manpower, it provides an anticipated understanding of the
properties of the galaxy distribution at large distances.

\acknowledgements

We are grateful to ESO for the numerous nights of observing time
allocated to this program. We also wish to thank the staff members
in La Silla who greatly contributed to the success of our observing
runs. This research is partly supported by the ``Programme National de
Cosmologie'' (previously ``GdR Cosmologie'') from INSU/CNRS.

\endacknowledgements

\newpage

\begin{table*}
  \label{tabccd}
  \caption{CCD photometric characteristics for EFOSC1 (3.6m) and EMMI (NTT)}
  \begin{tabular}{cccccccc}
  \hline
  Instrument & \multicolumn{3} {c} {Average zero-points} & \multicolumn{4} {c} {Color coefficients$^{(a)}$} \\
  CCD/Period                             &  B & V & R     
  & $k_B[B-V]$  &  $k_{V}[B-V]$ & $k_{V'}[V-R]$ & $k_R[V-R]$           \\
  \hline
  EFOSC1 & & & & & & & \\
  RCA\#8/43-44 & 23.70$\pm$0.03 & 24.24$\pm$0.04 & 24.21$\pm$0.02  &0.16$\pm$.03  & 0.04$\pm$.02& 0.10$\pm$.02 & 0.00$\pm$.02\\
  RCA\#8/45-46 & 23.41$\pm$0.01 & 24.06$\pm$0.03 & 24.03$\pm$0.02   &0.16$\pm$.03  & 0.04$\pm$.02& 0.10$\pm$.02 & 0.00$\pm$.02 \\
  \hline
  EMMI-R & & & & & & & \\ 
  THX\#18/49-50 &  & 24.30$\pm$0.02 & 24.65$\pm$0.03  &              & 0.05$\pm$.01& 0.10$\pm$.02& -0.10$\pm$.01 \\
  LOR\#34/52 & 24.69$\pm$0.02$^1$ &  & 25.14$\pm$0.02  &              &             &              & -0.03$\pm$.01 \\
  TEK\#36/54 &  & 25.40$\pm$0.02 &             &              & 0.03$\pm$.02& 0.05$\pm$.02&                \\
  \hline
  EMMI-B  & & & & & & & \\
  TEK\#31/52 &  24.69$\pm$0.02 & &  
        &-0.21$\pm$.02 &             &             &      \\
  TEK\#31/54 &  24.26$\pm$0.01 & &  
        &-0.21$\pm$.02 &             &             &             \\
  \hline
\end{tabular}
\smallskip
\\
\footnotesize
\underline{Note:} \\
$^{(a)}$ Indicated in brackets are the color terms by which must be
multiplied the listed coefficients $k_M$ in order to convert
the observed CCD magnitudes $M_{obs}$ into the standard
Johnson-Cousins magnitudes $M_{std}$ ($M_{std} = M_{obs} + k_M color$).\\
\end{table*}

{\bf Figure Captions:}

\begin{itemize}

\item {\it Figure 1}: Distribution on the sky of the 13,096 galaxies
in the ESO-Sculptor Survey (ESS) to its completeness limit of $R=23.5$
(in J2000 equatorial coordinates). The \rms magnitude uncertainties are
0.04 mag, and the \rms astrometric uncertainties are $\sim0.2$ arcsec. Even at
this large depth, the galaxy distribution shows large-sale
inhomogeneities which are measured by the angular 2-point correlation
function \cite{these-arnouts}.

\item {\it Figure 2:} 27 spectra from the ESS (black curves) and
their reconstructions using 3 principal components (red curves; see
text for details). Note the filtering effect on the spectra with low
signal-to-noise ratio, and the variable accuracy in the reconstruction
of spectral lines.

\item {\it Figure 3:} Distribution in R.A. versus redshift for 402
galaxies in the ESS in 3 contiguous redshift ranges (0.08--0.21,
0.21--0.34, 0.34--0.47). These maps show that the alternation of voids
and walls persist at large distances, with a typical scale of
20--60\hmpc \cite{ess-lss}.  The right-most cone shows the geometry of
the full volume of the ESS (430 galaxies). In all 4 cones, the right
ascension range is 4.72--5.9\deg (see Figure 1).

\item {\it Figure 4:} PCA spectral sequence for 277 ESS galaxies. The
parameter $\delta$ measures the relative contribution of the red and
blue stellar components in the galaxy (PC1 et PC2 in Figure 5), and
$\theta$ measures the contribution of the emission lines (as shown in
PC3 in Figure 5). Green points indicate galaxies with W[OII]$ \le
15$ \AA, blue points are galaxies with 15 \AA\ $\le$ W[OII]$ \le
30$ \AA, and red points are galaxies with W[OII]$ \ge 30$ \AA. The
different discrete classes obtained by comparison with the Kennicutt
spectra\cite{kenn} are indicated by vertical dotted lines. Note the
non-uniform sampling of the various types, with late-type galaxies
spanning a larger range in $\delta$, in good agreement the the larger
variations in morphological properties among the spiral galaxies. As
expected, late spectral types tend to have more frequent and stronger
emission lines. $\theta$ is a good indicator of emission line strength
for late-type galaxies when used in conjunction with $\delta$. Some of
the early-type objects also have emission lines \cite{ess-class}.

\item {\it Figure 5:} The first 3 principal components obtained for
the ESS sample. The 1st PC is the average spectrum and resembles an Sb
spectrum.  The 2nd PC allows to quantify the relative contribution of
the young stellar population, and the 3rd PC measures the contribution
from the emission lines. 98\% of the flux of the ESS spectra can be
reconstructed by linear combination of these 3 PCs \cite{ess-class}.

\item {\it Figure 6:} Observed and intrinsic $B-R$ colors for 330 ESS
galaxies as a function of redshift and spectral type (E/S0 in black;
Sa in red; Sb in green; Sc/Im in blue).  The differences result from
application of the K-corrections derived per spectral type, redshift
interval, and filter band, using the spectrophotometric model PEGASE
\cite{pegase}.

\item {\it Figure 7:} Luminosity functions in $B$ (open squares) 
and $R$ (filled circles) for 327 galaxies of the ESS. The best fit
Schechter functions have $M^*_B=-19.58\pm0.17$ $\alpha_B=-0.85\pm0.17$,
and $M^*_R=-21.15\pm0.19$ $\alpha_R=-1.23\pm0.13$ \cite{these-galaz}.

\end{itemize}


\begin{thebibliography}{99}

\bibitem{oort} Oort, J. H. 1983, \araa{21} 373
\bibitem{cfa-slice} de Lapparent, V., Geller, M. J., Huchra, J. P. 1986, 
      \apjl{302} L1
\bibitem{great-wall} Ramella, M., Geller, M. J., \& Huchra, J. P., 1992,
        \apj{384} 396
\bibitem{lcrs} Shectman, S. A., Landy, S. D., Oemler, A., Tucker,
   D. L., Kirshner, R. P., Lin, H., \& Schechter, P. L. 1996, \apj{470} 172 
\bibitem{esp} Vettolani, G., Zucca, E., Zamorani,  G., Cappi, A., 
  Merighi, R., Mignoli, M., Stirpe, G.M., MacGillivray, H., Collins, C.,
  Balkowski, C., Cayatte, V., Maurogordato, S., Proust, D., Chincarini, G.,
  Guzzo, L., Maccagni, D., Scaramella, R., Blanchard, A., Ramella, M. 1997,
  {\it Astron. \& Astroph.}, in press 
\bibitem{muenster} Schuecker, P. \& Ott, H.-A. 1991, \apjl{378} L1 (http://aquila.uni-muenster.de/mrsp-overview/mrsp-overview.html)
\bibitem{missing} Bahcall, N. A., Lubin, L. M., \& Dorman, V. 1995, \apjl{447} L81
\bibitem{nucleo} Dar, A. 1995, \apj{449} 553  
\bibitem{cdm} Blumenthal G. R., Faber, S. M., Primack, J. R., \& Rees, M. J. 
    1984, \nat{311} 517
\bibitem{biasing} Bardeen, J. M., Bond, J. R., Kaiser, N., Szalay, A. S 1986,
       \apj{304} 15
\bibitem{brosch} Brosch, N. 1989, \apj{344} 597
\bibitem{faint-void} Kuhn, B., Hopp, U., Elsaesser, H. 1997, \aa{318} 405
\bibitem{cobe} Bennett, C. L., Banday, A. J., Gorski, K. M., Hinshaw, G., 
      Jackson, P., Keegstra, P., Kogut, A., Wilkinson, D. T., Wright, E. L. 
      1996, \apjl{464} L1
\bibitem{mass-func} Ashman, K. M., Salucci, P., Persic, M. 1993, \mn{260} 610
\bibitem{madau} Madau, P., Ferguson, H. C., Dickinson, M. E., Giavalisco, M.,
        Steidel, C. C., Fruchter, A. 1996, \mn{283} 1388
\bibitem{segre} Santiago, B. X., da Costa, L. N. 1990, \apj{362} 386 
\bibitem{ess-phot} Arnouts, S., Lapparent, V., Mathez, G., Mazure, A.,
        \& Mellier, Y., Bertin, E., \& Kruszewski, A. 1997, \aas{124} 163
\bibitem{sex} Bertin, E. \& Arnouts, S. 1996, \aas{117} 393
\bibitem{these-arnouts} Arnouts, S., {\it Th\`ese de Doctorat}, 
   Universit\'e Paris VII, 1996.
\bibitem{met-nat} Metcalfe, N., Shanks, T., Campos, A., \& Fong, R. 1996, 
   \nat{383} 236
\bibitem{kenn} Kennicutt, R. C. 1992, \apjs{79} 255
\bibitem{ess-spec} Bellanger, C., de Lapparent, V., Arnouts, S.,
        Mathez, G., Mazure, A., \& Mellier, Y., 1995, \aas{110} 159
\bibitem{ess-class} Galaz, G., \& de Lapparent, V., 1997, 
   {\it Astron. \& Astroph.}, in press (astro-ph/9711093)
\bibitem{cfrs-tresse} Tresse, L., Rola, C., Hammer, F., Stasinska, G., 
   Le F\`evre, O., Lilly, S. J., Crampton, D. 1996, \mn{281} 847
\bibitem{ess-lss} Bellanger, C., de Lapparent, V., 1995, \apjl{455} L103
\bibitem{koo} Koo, D. C., Ellman, N., Kron, R. G., Munn, J. A., Szalay, A. S., 
        Broadhurst, T. J., \& Ellis, R. S., 1993, in ``Observational 
        Cosmology'', eds. G. Chincarini \etal, {\it ASP Conf. Ser.}, 
        Vol. {\bf 51}, 112
\bibitem{uv-morph} O'Connell, R. W., Marcum, P. 1996, in ``HST and the 
High Redshift Universe'' (37th Hertstmonceux Conference), eds. N.R. Tanvir, 
A. Aragon-Salamanca, \& J.V. Wall
\bibitem{murtagh} Murtagh, F., \& Heck, A. 1987, {\it Multivariate
Data Analysis}, Reidel
\bibitem{connolly} Connolly, A. J., Szalay, A. S., Bershady, M. A., 
      Kinney, A. L., \& Calzetty, D. 1995, \aj{110} 1071
\bibitem{sodre} Sodr\'e, L., \& Cuevas, H. 1994, {\it Vistas in Astronomy},
     {\bf 38}, 287
\bibitem{aaronson} Aaronson, M. 1978, \apj{221} L103
\bibitem{these-galaz} Galaz, G., {\it Th\`ese de Doctorat}, 
      Universit\'e Paris VII, 1997
\bibitem{dressler} Dressler, A. 1980, \apj{236} 351
\bibitem{pegase} ``Base de donn\'ees PEGASE'', 1996, A.A.S. CD-ROM Series, 
     {\bf Vol 7}, ed. Leitherer et al. 
     (http://www.iap.fr/users/rocca/index.html)
\bibitem{marzke-lf2} Marzke, R. O., Geller, M. J., Huchra, J. P., \&
        Corwin, H. G., Jr 1994, \aj,{108} 437 
\bibitem{bertin} Bertin, E., \& Dennefeld, M. 1997,  1997, \aa{317} 43
\bibitem{lcrs-lf} Lin, H., Kirshner, R. P., Shectman, S. A., Landy, S. D., 
        Oemler, A., Tucker, D. L., \& Schechter, P. L. 1996, \apj{464}
        60
\bibitem{esp-lf} Zucca, E., Zamorani,  G., Vettolani, G., Cappi, A., 
  Merighi, R., Mignoli, M., Stirpe, G.M., MacGillivray, H., Collins, C.,
  Balkowski, C., Cayatte, V., Maurogordato, S., Proust, D., Chincarini, G.,
  Guzzo, L., Maccagni, D., Scaramella, R., Blanchard, A., Ramella, M. 1997,
  {\it Astron. \& Astroph.}, in press 
\bibitem{autofib-lf2} Heyl, J., Colless, M., Ellis, R. S., \& Broadhurst, T.,
        1997, {\it M{.}N{.}R{.}A{.}S{.}}, in press (astro-ph/9610036)
\bibitem{cnoc1-lf} Lin, H., Yee, H. K. C., Carlberg, R. G., Ellingson, E. 1997,
        \apjl{475} 494
\bibitem{cfrs-lf} Lilly, S. J., Tresse, L., Hammer, F., Crampton, D., \& 
        Le F\`evre, O., 1995, \apj{455} 108
\bibitem{hdf} Sawicki, M. J., Lin, H., Yee, H. K. C. 1997, \aj{113} 1
\bibitem{bo} Butcher, H. \& Oemler, A. 1978, \apj{219} 18
\bibitem{cowie} Cowie, L. L., Gardner, J. P., Hu, E. M., Songaila, A.,
         Hodapp, K. -W., Wainscoat, R. J. 1994, \apj{434} 114
\bibitem{white} White, S. D. M. 1997, in ``The Early Universe with the VLT'',
    ed. J. Bergeron (Springer-Verlag)
\bibitem{sdss} http://www-sdss.fnal.gov:8000/
\bibitem{2dF} http://msowww.anu.edu.au/~colless/2dF/
\bibitem{deep} http://www.ucolick.org/~deep/home.html
\bibitem{lzt} http://www.astro.ubc.ca/LMT/lzt.html

\end{thebibliography}
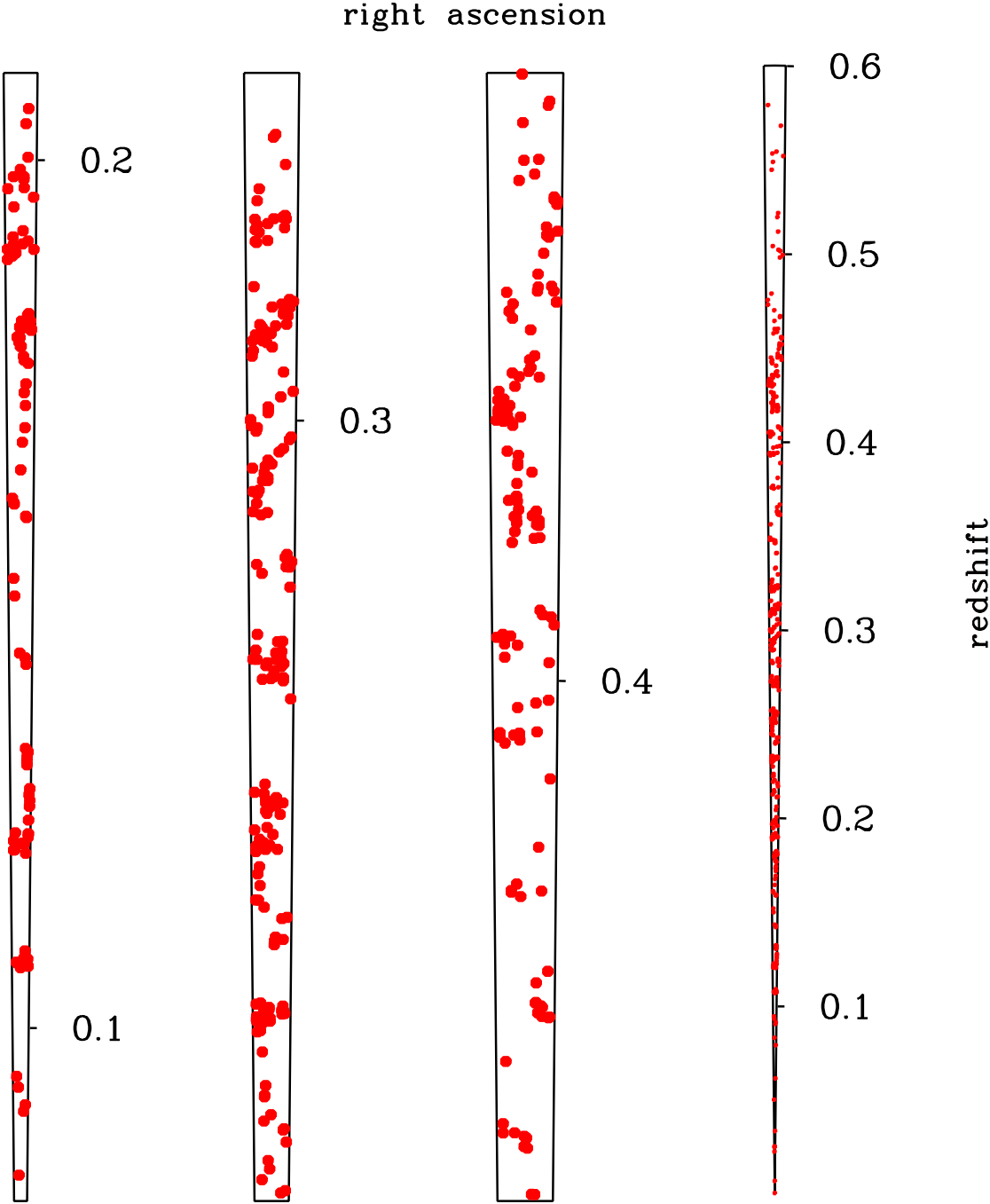
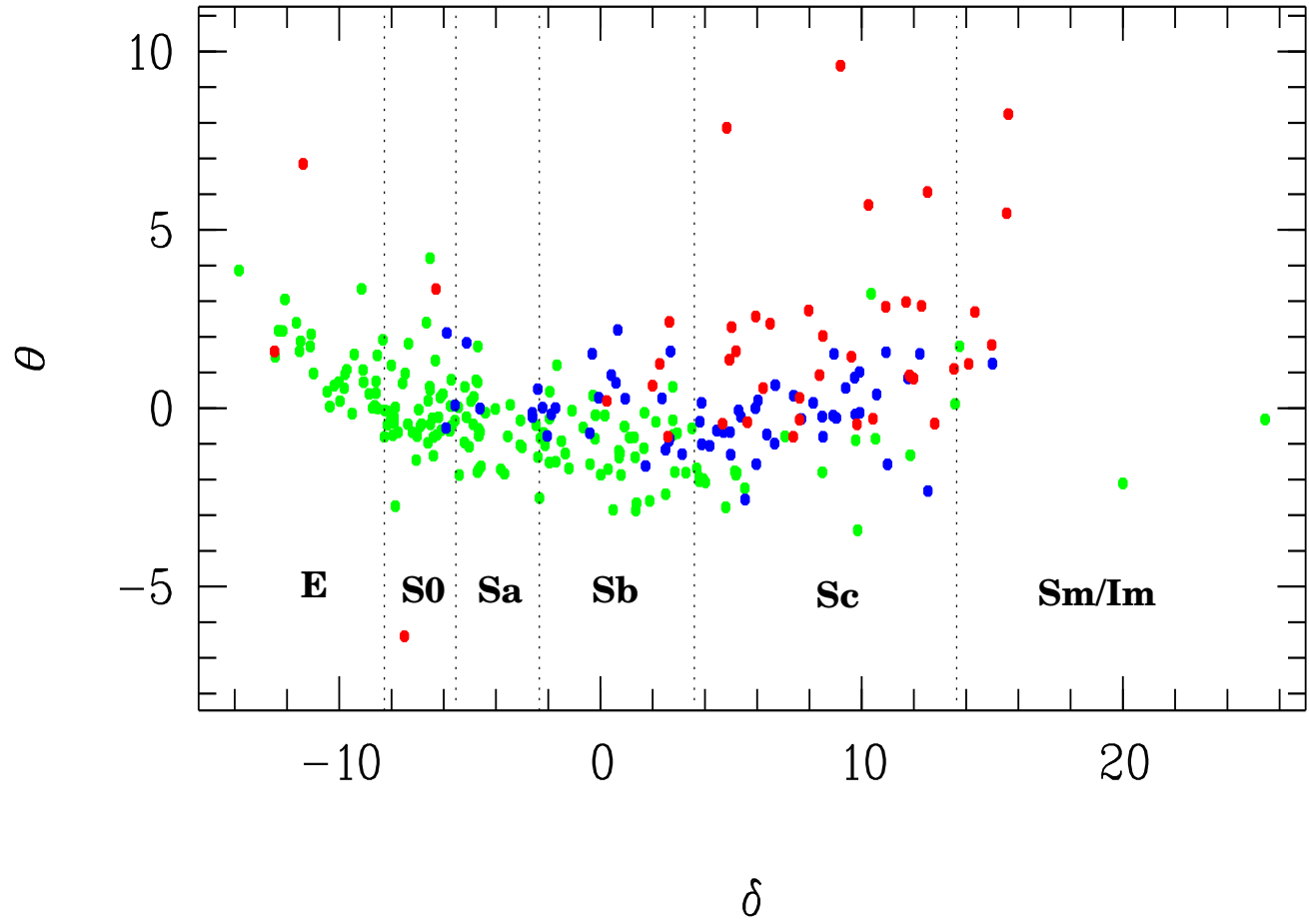
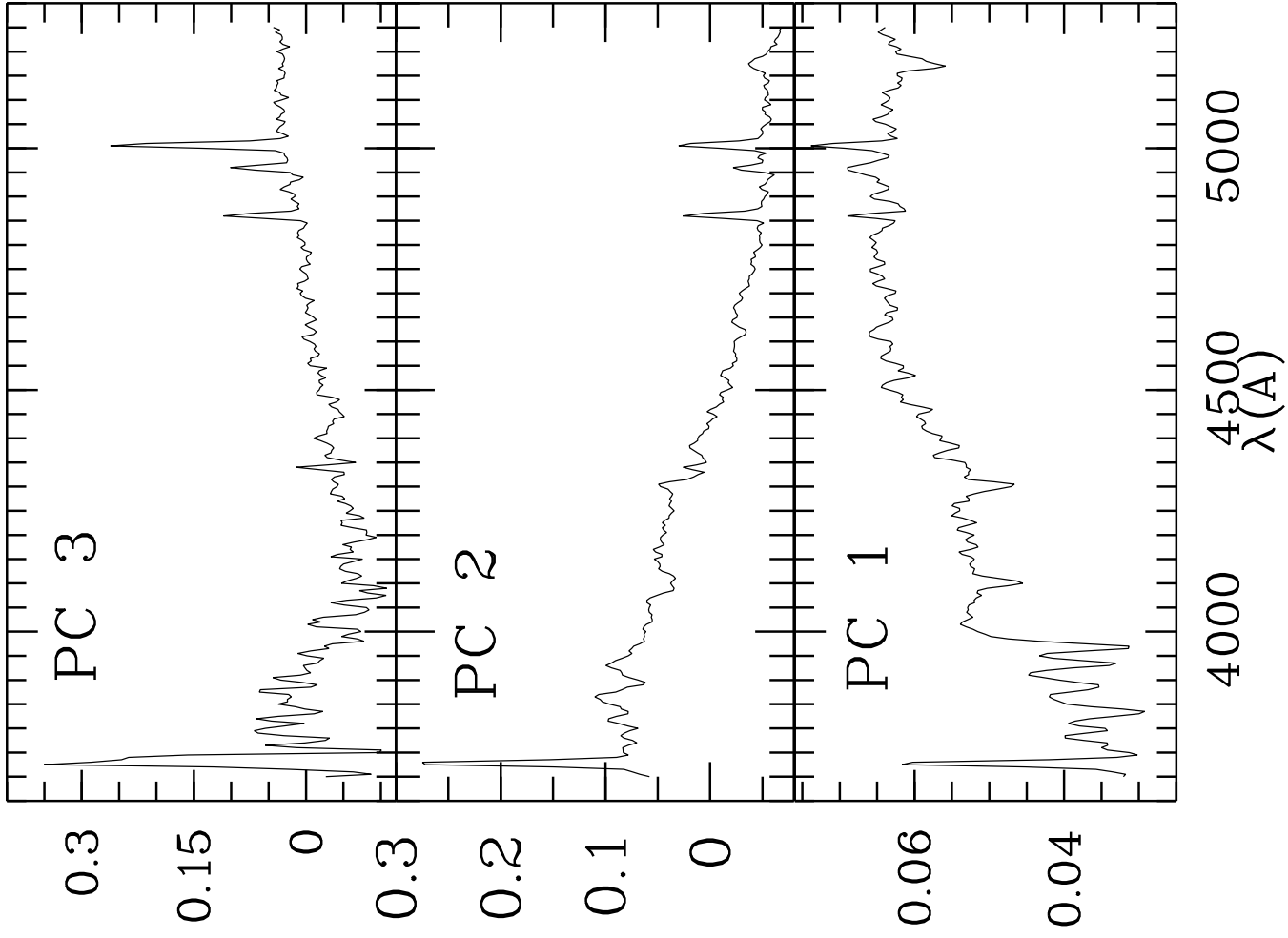
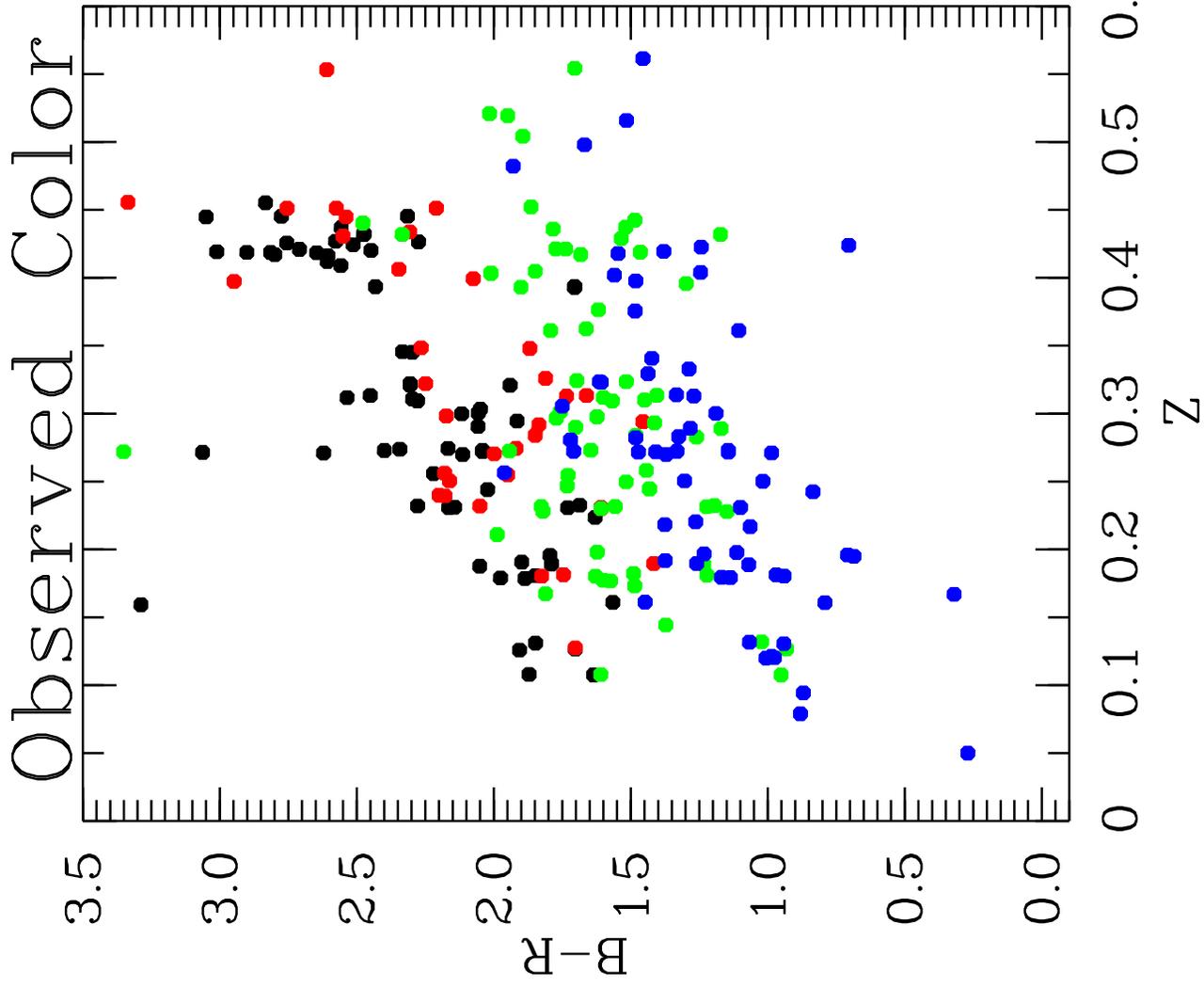
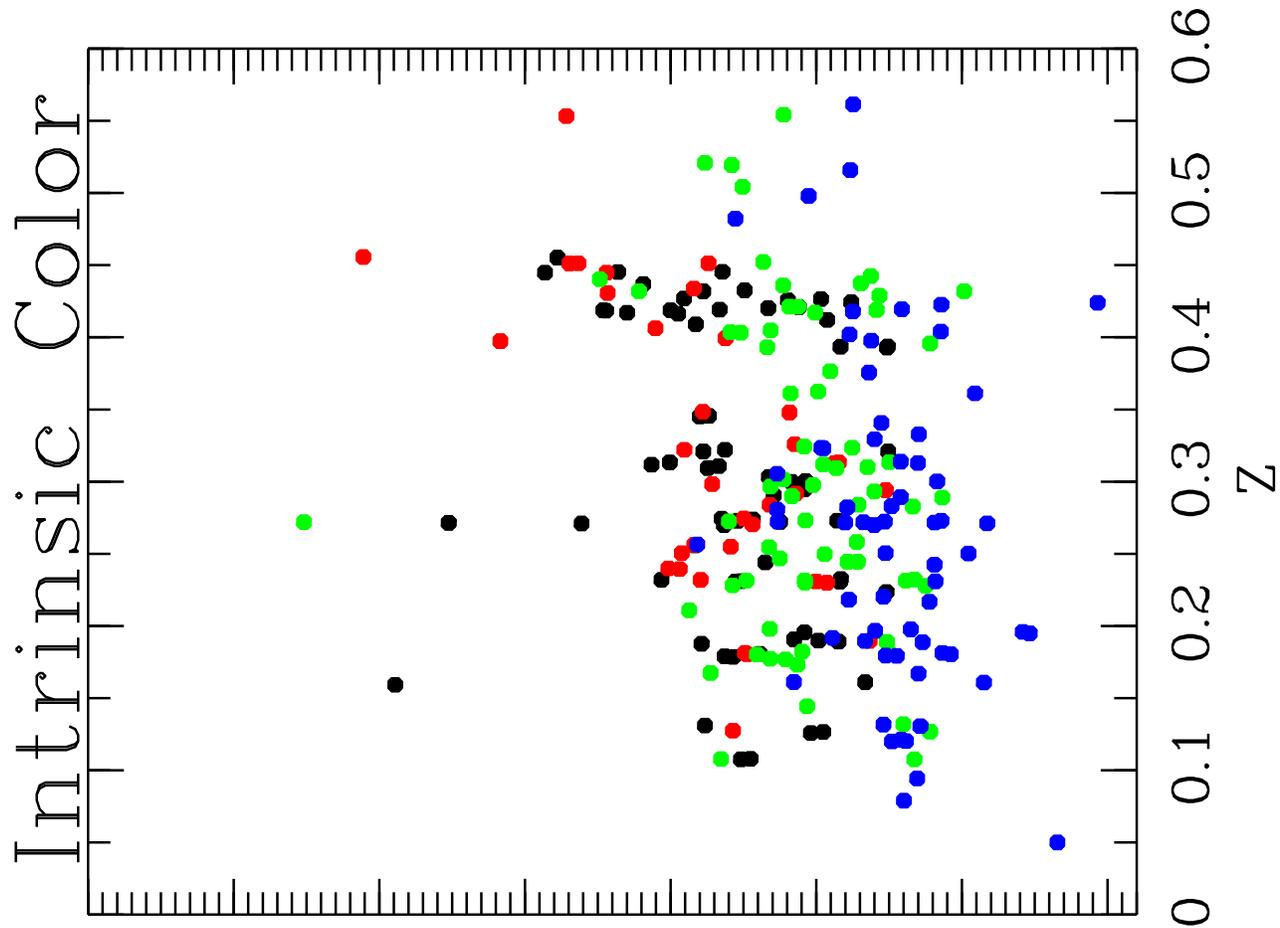
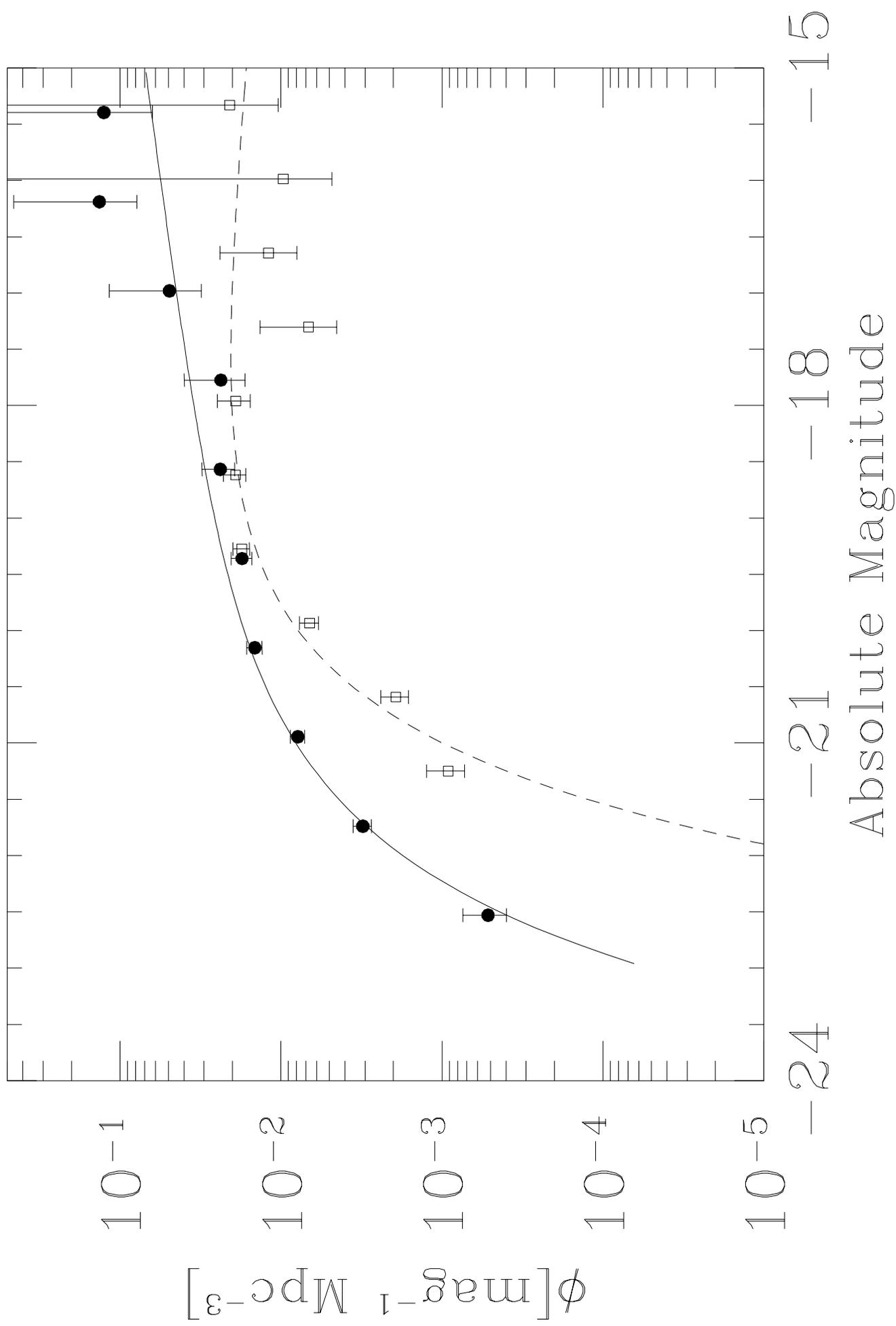
\end{document}